\documentclass[conference]{IEEEtran}
\usepackage{color}
\usepackage{balance}
\usepackage{graphicx}
\usepackage{enumerate}
\usepackage{subfigure}
\usepackage{algorithm}
\usepackage{algorithmic}
\usepackage{booktabs}
\usepackage{cite}
\usepackage{amssymb}
\usepackage{amsmath}
\usepackage{float}
\usepackage{float}
\usepackage{bm} 

\usepackage{fancyhdr}
\usepackage{url}

\begin{document}

\title{ \huge Intelligent Reflecting Surface Assisted Anti-Jamming Communications Based on Reinforcement Learning 
\vspace{5pt}}

 \author{Helin Yang$^1$, Zehui Xiong$^1$, Jun~Zhao$^1$, Dusit Niyato$^1$, Qingqing Wu$^2$,  Massimo Tornatore$^3$, and Stefano Secci$^4$ \\
$^1$School of Computer Science and Engineering, Nanyang Technological University, Singapore\\
$^2$State Key Laboratory of Internet of Things for Smart City, University of Macau, Macau, 999078 China\\
$^3$Department of Electronics,  Information and Bioengineering, Politecnico di Milano,  Milan, Italy\\
 $^4$ Cnam, Cedric Lab, Paris, France\\
E-mail: \{hyang013, zxiong002, junzhao,  dniyato\}@ntu.edu.sg
\vspace{-20pt} 
}

\maketitle

 \thispagestyle{fancy}
\pagestyle{fancy}
\lhead{ \footnotesize This paper appears in the Proceedings of IEEE Global Communications Conference (\textbf{GLOBECOM}) 2020. \hfill \footnotesize \url{http://JunZhao.info/} \\[3pt]  A full version appears in \normalsize \textbf{IEEE Transactions on Wireless Communications}.  \hfill \footnotesize \url{https://arxiv.org/abs/2004.12539} \\[3pt] $~~$~\hfill \footnotesize \url{https://doi.org/10.1109/TWC.2020.3037767}}
\cfoot{\thepage}
\renewcommand{\headrulewidth}{0.4pt}
\renewcommand{\footrulewidth}{0pt}

 \linespread{1.00}{

\begin{abstract}

Malicious jamming launched by smart jammer, which  attacks legitimate transmissions has been  regarded as one of the critical security challenges in wireless communications. Thus,  this paper exploits intelligent reflecting surface (IRS) to enhance anti-jamming communication performance and mitigate jamming interference by adjusting the surface reflecting elements at the IRS. Aiming to enhance the communication performance against smart jammer, an optimization problem for jointly optimizing power allocation at the base station (BS) and reflecting beamforming at the IRS is formulated. As the jamming model and jamming behavior are dynamic and unknown, a win or learn fast policy hill-climbing (WoLF–PHC) learning approach is proposed to jointly optimize the anti-jamming power allocation and reflecting beamforming strategy without the knowledge of the jamming model. Simulation results demonstrate that the proposed anti-jamming based-learning approach can efficiently improve both  the IRS-assisted system rate and transmission protection level compared with existing solutions.  \vspace{5pt}

\textbf{\emph{Index Terms}}---Anti-jamming, intelligent reflecting surface, power allocation, beamforming, reinforcement learning.
\end{abstract}
}
 
\vspace{-7pt} 

\IEEEpeerreviewmaketitle
\section{Introduction}
\IEEEPARstart{D}{ue} to the inherent broadcast and openness nature of wireless channels [1], [2], wireless transmissions can be easily vulnerable to jamming attacks. In particular, malicious jammers can intentionally send jamming signals over the legitimate channels to degrade communication performance [1]-[3], which has been considered as  one of the serious threats in wireless communications.  In this regard, lots of jamming-related studies  have been recently presented to defend jamming attacks, including  frequency hopping, power control, relay assistance, beamforming, and so on.

Frequency-hopping (FH) is one of the  powerful techniques which has been widely  adopted to allow a wireless user to quickly switch its current operating frequency to other frequency spectrum, thereby avoiding potential  jamming attacks [4]-[6].  In [4] and [5], a mode-FH approach was presented to jointly utilize conventional FH to further improve the communication performance in the presence of jammers.  In [6], Hanawal $et~al$. proposed a joint FH and rate adaptation scheme to avoid jamming attacks in the presence of a jammer.  Besides FH, power control is another commonly used technique, e.g., [3], [7]-[9]. As an example, [7] and [8] investigated a jammed wireless system where the system operator tries to control the transmit power to maximize system rate. The authors in [3] and [9] studied the anti-jamming problem with power control strategies, by leveraging the  game theory to optimize the power control policy of the transmitter against jammers. Moreover, cooperative communication using trusted relays has been proposed as one promising anti-jamming technique for improving the physical layer security [10]-[12], and robust joint cooperative beamforming and jamming designs  were proposed to maximize the achievable rate under the imperfect channel state information (CSI) of a jammer.

To deal with uncertain and/or  unknown jamming attack models, such as jamming policies and jamming power levels, some existing studies utilized reinforcement learning (RL) algorithms have been applied in some existing studies  to optimize the jamming resistance policy in dynamic wireless communication systems [13]-[15]. In [13], a policy hill climbing (PHC)-based Q-Learning approach was studied to improve the communication performance against  jamming without knowing the jamming model. In [14] and [15], the authors adopted deep reinforcement learning (DRL) algorithms that enable transmitters to quickly obtain an optimal policy to guarantee security performance against jamming.
     
However, despite the effectiveness of the above mentioned anti-jamming schemes [3]-[15], employing  a  number  of  active  relays  incurs an  excessive  hardware  cost, and anti-jamming beamforming and power control in communication systems is generally  energy-consuming. To tackle these shortcomings, a new paradigm, called intelligent reflecting surface (IRS) [16], [17], has been recently proposed as a promising technique to enhance the secrecy performance. In particular, IRS   comprises of a large number of low-cost passive reflecting elements, where each of the elements adaptively adjusts its reflection amplitude and/or phase to control the strength and direction of the reflected electromagnetic wave [16], [17]. As a result, IRS has been employed in wireless commutation systems to devote to security performance optimization [18]-[22]. In [18]-[21], the authors investigated the physical layer security enhancement of IRS-assisted communications systems, where both the BS’s beamforming and the IRS’s phase shifts were jointly optimized to improve secrecy rate in the presence of an eavesdropper. Furthermore, Yang $et~al$. in [22] applied DRL to learn the secure beamforming policy in multi-user IRS-aided secure systems, in order to maximize the system secrecy rate in the presence of multiple  eavesdroppers. To the best of our knowledge, IRS has not been explored yet in the existing works [3]-[22] to enhance the anti-jamming strategy against smart jamming.

In this paper, we propose an IRS-assisted anti-jamming solution for  securing  wireless communications. In particular, we aim to maximize the system rate of multiple legitimate users in the presence of a smart multi-antenna jammer. As the jamming model and jamming behavior are dynamic and unknown, a  win or learn fast policy hill-climbing (WoLF–PHC) anti-jamming approach is proposed to achieve the optimal anti-jamming  policy, where WoLF–PHC is capable of quickly achieving the optimal policy without knowing the jamming model. Simulation results verify the effectiveness of the proposed learning approach in terms of improving the system rate, compared with the existing approaches.

The remainder of this paper is organized as follows. Section II provides the system model and problem formulation. Section III proposes  the  WoLF–PHC-based learning approach. Simulation results are provided in Section IV, and the paper is concluded in Section V.

\section{System Model and Problem Formulation}

\subsection{System Model}
 
\begin{figure}
\centering

\includegraphics[width=0.85\columnwidth]{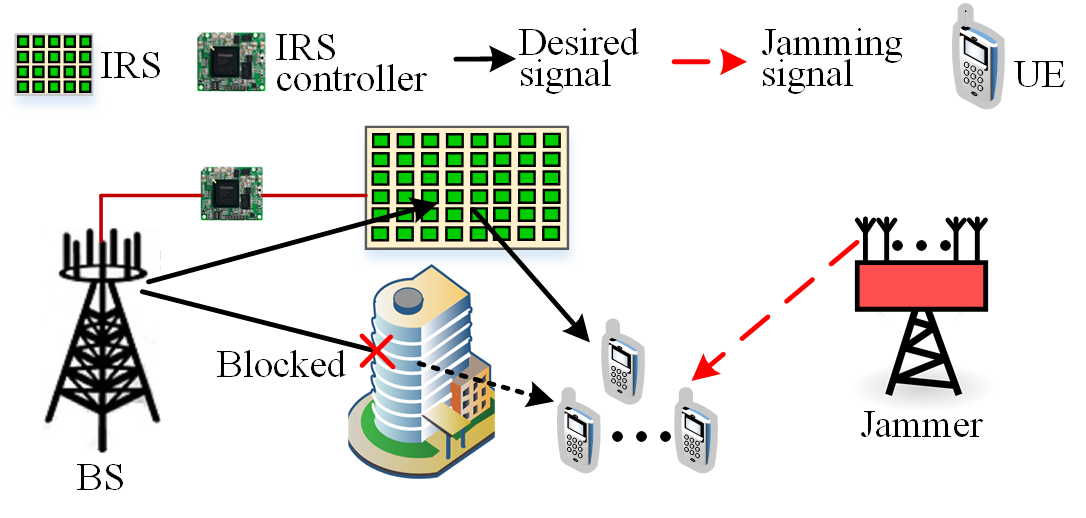}
\caption{{Illustration of an IRS-assisted communication system against a multi-antenna jammer.} } \label{fig:Schematic}

\end{figure}

As shown in Fig. 1, this paper considers an IRS-assisted communication system, which
consists of one BS with $N$ antennas and $K$ single-antenna
legitimate user equipments (UEs) located at the cell-edge.
The IRS comprised of $M$ reflecting elements is deployed to provide additional communication links so
as to improve the performance for the UEs over a given frequency
band. The direct communication links of cell-edge UEs may suffer high signal attenuation and
these links are severely blocked
by obstacles when these UEs located in dead zones. In addition, as
illustrated in Fig. 1,  a malicious multi-antenna jammer is located near the
legitimate UEs who attempts to interfere the legitimate
transmissions by sending faked or replayed jamming signal for the
UEs via ${N_{\rm{J}}}$ antennas, in order to degrade the legitimate
communication performance. In this case, deploying the IRS can
effectively enhance the desired signal power and mitigate the
jamming interference generated from the jammer by designing the
reflecting beamforming at the IRS.

Let  $\mathcal{K} = \{ 1,2,...,K\} $ and $\mathcal{M} = \{
1,2,...,M\} $ represent the UE set and the IRS reflecting element
set, respectively.  Let ${\bf{G}} \in {{\mathbb{C}^{M \times N}}}$,
${\bf{g}}_{{\rm{bu,}}k}^H \in {\mathbb{C}^{1 \times N}}$,
${\bf{g}}_{{\rm{ru,}}k}^H \in {\mathbb{C}^{1 \times M}}$, and
${\bf{h}}_{{\rm{J,}}k}^H \in {\mathbb{C}^{1 \times {N_{\rm{J}}}}}$
denote the channel coefficients between the BS and the IRS, between the
BS and the $k$-th UE, between the IRS and the $k$-th UE, and between the
jammer and the $k$-th UE, respectively. The quasi-static flat-fading model is assumed for all the above channels. Let ${\bf{\Phi }} = {\rm{diag}}({\Phi
_1},{\Phi _2},....,{\Phi _M}) \in {\mathbb{C}^{M \times M}}$ denotes the
reflection coefficient matrix associated with effective phase
shifts at the IRS, where ${\Phi _m} = {\omega _m}{e^{j{\theta
_m}}}$ comprises both  an reflection amplitude   ${\omega _m} \in [0,1]$ and
a phase shift coefficient  ${\theta _m} \in [0,2\pi ]$ on the
combined received signal. Since each phase shift is
favorable to be designed to achieve  maximum signal reflection,
we consider that ${\omega _m} = 1$, $\forall m \in \mathcal{M}$ in this paper [16]-[22].

The transmitted signal at the BS can be expressed as 
\begin{equation}
\begin{split}
{\bf{x}} = \sum\nolimits_{k = 1}^K {\sqrt {{P_k}}
{{\bf{w}}_k}{s_k}}
\end{split}
\end{equation}
where ${P_k}$ stands for the transmit power allocated for the $k$-th
UE and we have the power constraint: $\sum\nolimits_{k =
1}^K {{P_k}}  \le {P_{\max }}$ with ${P_{\max }}$ being the
maximum transmit power of the BS, ${s_k}$ is the transmitted symbol for the
$k$-th UE, ${s_k} \in \mathbb{C}$, $\mathbb{E}\{ {s_k}\}  = 0$ and
$\mathbb{E}{\rm{\{ |}}{s_k}{{\rm{|}}^{\rm{2}}}{\rm{\} }} =
{\rm{1}}$ which denotes the unit power information symbol, and
${{\bf{w}}_k} \in {\mathbb{C}^{N \times 1}}$ is the beamforming vector for
the $k$-th UE with ${\left\| {{{\bf{w}}_k}} \right\|^2} = 1$,
respectively.

This paper considers the case that the smart jammer attempts to
disturb the BS's transmitted signal by emitting jamming signal
${{\bf{z}}_k} \in {\mathbb{C}^{{N_{\rm{J}}} \times 1}}$ to attack the $k$-th UE.  In addition, the
transmit power of the faked jamming signal for the $k$-th UE is
denoted as ${P_{{\rm{J}},k}} = {\left\| {{{\bf{z}}_k}} \right\|^2} =
{\rm{Tr}}({{\bf{z}}_k}{\bf{z}}_k^H)$. In this case, for 
UE $k$, the received signal consists of the signal coming from its
associated BS, the reflected signal from the IRS and the jamming
signal from the jammer, which is written by
\begin{equation}
\begin{split}
\begin{array}{l}
\noindent {y_k} = \underbrace {\left( {{\bf{g}}_{{\rm{ru}},k}^H{\bf{\Phi }}{\bf{G}}
+ {\bf{g}}_{{\rm{bu}},k}^H} \right)\sqrt {{P_k}}
{{\bf{w}}_k}{s_k}}_{{\rm{desired}}\;{\rm{signal}}} + \\
\underbrace
{\sum\limits_{i \in \mathcal{K},i \ne k} {\left(
{{\bf{g}}_{{\rm{ru}},k}^H{\bf{\Phi }}{\bf{G}} + {\bf{g}}_{{\rm{bu}},k}^H}
\right)\sqrt {{P_i}} {{\bf{w}}_i}{s_i}} }_{{\rm{inter-user
~interference}}} + \underbrace {\sqrt {{P_{{\rm{J}},k}}}
{\bf{h}}_{{\rm{J}},k}^H{{\bf{z}}_k}}_{{\rm{jamming}}\;{\rm{signal}}}
+ {n_k}
\end{array}
\end{split}
\end{equation}
where  ${n_k}$ denotes the additive complex Gaussian noise with the
zero mean and variance $\delta _k^2$  at the $k$-th UE. In (2), in
addition to the received desired signal, each UE also suffers
inter-user interference (IUI) and the jamming interference signal
in the system. According to (2), the received signal-to-interference-plus-noise-ratio (SINR) at the $k$-th UE can be expressed as
\begin{equation}
  { 
SIN{R_k} = \frac{{{P_k}{{\left| {\left(
{{\bf{g}}_{{\rm{ru}},k}^H{\bf{\Phi }}{\bf{G}} + {\bf{g}}_{{\rm{bu}},k}^H}
\right){{\bf{w}}_k}} \right|}^2}}}{{\sum\limits_{i \in \mathcal{K},i \ne k}
{{P_i}{{\left| {\left( {{\bf{g}}_{{\rm{ru}},k}^H{\bf{\Phi }}{\bf{G}} +
{\bf{g}}_{{\rm{bu}},k}^H} \right){{\bf{w}}_i}} \right|}^2}}  +
{P_{{\rm{J}},k}}{{\left| {{\bf{h}}_{{\rm{J}},k}^H{{\bf{z}}_k}}
\right|}^2} + \delta _k^2}}. }
\end{equation}

\subsection{Problem Formulation}

In this paper, we aim to jointly optimize the transmit power
allocation  ${\{ {P_k}\} _{k \in \mathcal{K}}}$ at the BS and the reflecting
beamforming matrix ${\bf{\Phi }}$ at the IRS to maximize the
system achievable rate of all UEs against smart jamming,
subject to the transmit power constraint. Accordingly, the optimization problem can be formulated as
\begin{equation}
\begin{split}
\begin{array}{l}
\mathop {\max }\limits_{{{\{ {P_k}\} }_{k \in \mathcal{K}}},{\bf{\Phi }}} \; \sum\limits_{k \in \mathcal{K}} {{{\log }_2}} \left( {1 + SIN{R_k}} \right)\\
s.t.\;({\rm{a}}):\;\sum\nolimits_{k = 1}^K {{P_k}}  \le {P_{\max }},\\
\;\;\;\;\;\;({\rm{b}}):\;|{\Phi _m}| = 1,\;0 \le {\theta _m} \le 2\pi
,\;\forall m \in \mathcal{M}
\end{array}
\end{split}
\end{equation}
Note that  problem (4) is a non-convex optimization
problem, where the objective function is non-concave over the reflecting
beamforming matrix
${\bf{\Phi }}$; furthermore, the transmit power
allocation  variables  ${\{ {P_k}\} _{k \in
\mathcal{K}}}$ and ${\bf{\Phi }}$
  are intricately coupled in the objective function, thus rendering the joint optimization problem difficult to be solved optimally. So far, many optimization algorithms
[16]-[21] have been proposed to obtain  an approximate
solution to problem  (4), by iteratively updating either
${\{ {P_k}\} _{k \in
\mathcal{K}}}$ or ${\bf{\Phi }}$   with the other fixed at each iteration. Hence, this paper proposes an
effective solution  to address such kind of the optimization problem, which will be provided in the next section. In addition, it is worth noting that this paper mainly pays attention to jointly optimize the power allocation and the reflecting beamforming, so the transmit beamforming vector ${{\bf{w}}_k}$ is set by maximizing the received signal power at the IRS as the directin link from the BS to the UEs  suffer high signal attenuation by obstacles [16], [17].

\section{Joint Power Allocation and Reflecting Beamforming Based on RL}

The problem formulated in (4) is difficult to be solved as
mentioned at the end of the last section. Model-free RL is one of the dynamic programming tools which has the ability
to address the decision-making problem by achieving an optimal
policy in dynamic uncertain environments [33]. Thus, this paper
models the optimization problem as an RL, and a 
WoLF-PHC-based joint power allocation and reflecting beamforming
approach is proposed to learn the optimal anti-jamming strategy. 

\subsection{Optimization Problem Transformation Based on RL}

In RL, the IRS-assisted communication system is acted as an
environment and the central controller at the BS is regarded as a
 learning agent. In addition to the environment and the agent,
an RL also includes a set of possible system states $\mathcal{S}$,
a set of available actions $\mathcal{A}$, and a reward function
$r$, where the learning agent continually learns by interacting with the environment. The main elements
of RL are introduced as follows:

\textbf{States:} The system state ${s^t} \in \mathcal{S}$  is the
discretization of the observed information from the environment at
the current time slot $t$. The system state ${s^t}$ includes the
previous jamming power, i.e., ${\{ P_{{\rm{J}},k}^{t - 1}{\rm{\}
}}_{k \in \mathcal{K}}}$ according the channel quality, the
previous UEs' SINR values ${\{ SINR_k^{t - 1}{\rm{\} }}_{k \in
\mathcal{K}}}$, as well as the current estimated channel
coefficients ${\{{\bf{g}}_k^t{\rm{\} }}_{k \in \mathcal{K}}}$,
which is defined as
\begin{equation}
\begin{split}
{s^t} = \left\{ {{{\{ P_{{\rm{J}},k}^{t - 1}{\rm{\} }}}_{k \in
\mathcal{K}}},{{\{{\bf{g}}_k^t{\rm{\} }}}_{k \in
\mathcal{K}}},{{\{ SINR_k^{t - 1}{\rm{\} }}}_{k \in \mathcal{K}}}}
\right\}.
\end{split}
\end{equation}

\textbf{Actions:}  The action  ${a^t} \in \mathcal{A}$ is one of
the valid selections that the learning agent chooses at the time
slot $t$, which includes the transmit power  ${\{ {P_k}\} _{k \in
\mathcal{K}}}$ and the reflecting beamforming
coefficient (phase shift)  ${\{ {\theta _m}\} _{m \in
\mathcal{M}}}$. Hence, the action  ${a^t}$ is given by
\begin{equation}
\begin{split}
{a^t} = \left\{ {{{\{ P_k^t\} }_{k \in \mathcal{K}}},{{\{ \theta
_m^t\} }_{m \in \mathcal{M}}}} \right\}.
\end{split}
\end{equation}

\textbf{Transition probability:} $\mathcal{P}( \cdot )$ is a
transition model which represents the probability of taking an action
$a$ at a current state $s$ and then ending up in the next state
$s'$, i.e., $\mathcal{P}(s'|s,a)$.

\textbf{Policy:} Let $\pi ( \cdot )$ denotes a policy and it maps
the current system state to a probability distribution over the
available actions which is taken by the agent, i.e., $\pi (a,s):S
\to \mathcal{A}$.

\textbf{Reward function:} The reward function design plays an
important role in the policy learning in RL, where the reward
signal correlates with the desired goal of the system performance.
In the  optimization problem considered in Section
II.B, our objectives are twofold: maximizing the UEs' achievable
rate while decreasing the power consumption at the BS as much  as
possible. 

Based on the above analysis, the reward function is set as
\begin{equation}
\begin{split}
r = \underbrace {\sum\limits_{k \in \mathcal{K}} {{{\log }_2}}
\left( {1 + SIN{R_k}} \right)}_{{\rm{part}}\;{\rm{1}}} -
\underbrace {{\lambda _1}\sum\limits_{k \in \mathcal{K}} {{P_k}}
}_{{\rm{part}}\;2}
\end{split}
\end{equation}

In (7), the part 1 represents the immediate utility (system
achievable rate), the part 2 is the cost functions
which is defined as the transmission cost of the power
consumption at the BS, with ${\lambda _1}$ being the corresponding coefficient.

\subsection{WoLF-PHC-Based Joint Power Allocation and Reflecting
Beamforming}

Most of existing RL algorithms are value-based RL, such as
Q-Learning, Deep Q-Network (DQN) and double DQN. These 
RL algorithms can estimate the Q-function with low variance as well as
adequate exploration of action space, which can be ensured by
using the greedy scheme. In addition, policy gradient based RL
algorithm has the ability to tackle the continuous action space
optimization problems, but it may converge to suboptimal solutions
[22].

\begin{figure}
\centering
\includegraphics[width=0.85\columnwidth]{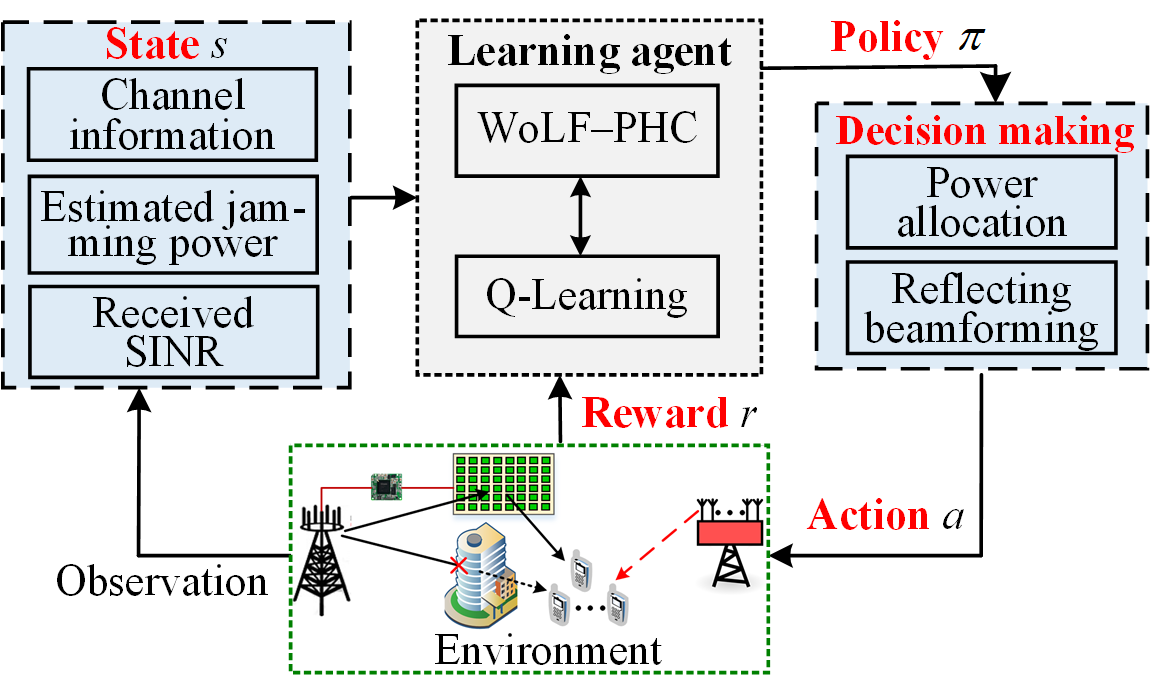}
\caption{{WoLF-PHC-based anti-jamming policy for
IRS-assisted systems.} } \label{fig:Schematic}
\end{figure}

In order to obtain the optimal anti-jamming policy against smart
jamming, we propose a fast WoLF-PHC-based joint power
allocation and reflecting beamforming for IRS-assisted
communication systems, as shown in Fig. 2, where WoLF-PHC is
utilized to enable the learning agent to learn and adapt faster in
dynamic uncertain environments. In the IRS-assisted system, the learning agent observes
a system state and receives an instantaneous reward by interacting
with the environment. Then, such information is leveraged  to train the
learning model to choose the anti-jamming policy with the maximum
Q-function value. After that, according to the selected policy,
the action is chosen to make decision in terms of power allocation
and reflecting beamforming. 

The objective of the learning agent is to obtain an optimal policy
that optimizes the long-term cumulative discounted reward instead
of its immediate reward, which can be expressed as ${R_t} =
\sum\nolimits_{j = 0}^\infty  {{\gamma ^j}{r^{(t + j + 1)}}} $,
where $\gamma  \in (0,1]$ denotes the discount factor. Adopting
${Q^\pi }({s^t},{a^t})$ as the state-action value function, which
represents the value of executing an action $a$ in a state $s$
under a policy $\pi $,  it can be expressed as
\begin{equation}
\begin{split}
{Q^\pi }({s^t},{a^t}) = {E_\pi }\left[ {\sum\limits_{j = 0}^\infty
{{\gamma ^j}{r^{(t + j + 1)}}} |{s^t} = s,{a^t} = a} \right].
\end{split}
\end{equation}

Similar to [23], the state-action  Q-function
${Q^\pi }({s^t},{a^t})$ satisfies the Bellman equation which is  expressed
as
\begin{align}  
{Q^\pi }({s^t},{a^t}) = {E_\pi }[  {r^{t + 1}} + \gamma
\sum\limits_{{s^{t + 1}} \in \mathcal{S}}  P({s^{t +
1}}|{s^t},{a^t})\\
\times \sum\limits_{{a^{t + 1}} \in \mathcal{A}} {\pi ({s^{t +
1}},{a^{t + 1}}){Q^\pi }({s^{t + 1}},{a^{t + 1}})}     ]. 
\end{align}

The conventional Q-Learning algorithm is widely utilzied to search
the optimal policy ${\pi ^ * }$. From (9), the optimal Q-function
(Bellman optimality equation) associated with the optimal policy
has the following form
\begin{equation} 
{Q^ * }({s^t},{a^t}) = {r^{t + 1}} + \gamma \sum\limits_{{s^{t +
1}} \in \mathcal{S}} {\mathcal{P}({s^{t + 1}}|{s^t},{a^t})} \mathop {\max
}\limits_{{a^{t + 1}} \in \mathcal{A}} {Q^*}({s^{t + 1}},{a^{t + 1}}). 
\end{equation}

It is worth noting that the Bellman optimality equation generally
does not have any closed-form solution. Thus, the optimal
Q-function (10) can be solved recursively to achieve the optimal
${Q^ * }({s^t},{a^t})$ by using an iterative method. Accordingly,
the updating on the state-action value function $Q({s^t},{a^t})$
is expressed as
\begin{equation}
\begin{split}
\begin{array}{l}
Q({s^t},{a^t}) \leftarrow (1 - \alpha )Q({s^t},{a^t})\\
\;\;\;\;\;\;\;\;\;\;\;\;\; + \alpha \left( {{r^t} + \gamma \mathop
{\max }\limits_{{a^t} \in \mathcal{A}} {Q^*}({s^{t + 1}},{a^t})} \right)
\end{array}
\end{split}
\end{equation}
where   $\alpha  \in (0,1]$ stands for the learning rate for the update of
Q-function.

The $\varepsilon-$greedy policy is capable of balancing the tradeoff between an exploitation and an
exploration in an RL, in order to avoid
converging to local optimal power allocation and reflecting
beamforming strategy. In the $\varepsilon  - $greedy policy, the
agent selects the action with the maximum Q-table value with
probability  ${\rm{1}} - \varepsilon $, whereas a random action is
picked with probability  $\varepsilon $ to avoid achieving stuck
at non-optimal policies [23]. Hence, the action selection
probability of the learning agent is expressed as 
\begin{equation}
\begin{split}
{\rm{Pr}}(a = \tilde a) = \left\{ \begin{array}{l}
{\rm{1}} - \varepsilon ,\;\;\;\;\tilde a = \arg \mathop {\max }\limits_{a \in \mathcal{A}} Q(s,a),\\
\frac{\varepsilon }{{|\mathcal{A}| - 1}},\tilde a \ne \arg \mathop {\max
}\limits_{a \in \mathcal{A}} Q(s,a).
\end{array} \right.
\end{split}
\end{equation}

As the WoLF-PHC algorithm is capable of not only keeping the
Q-function but also quickly learning the decision-making policy under uncertain characteristics [24], so this paper adopts it to derive the optical power allocation and reflecting beamforming
strategy with the unknown jamming model.
In the IRS-assisted communication system, the WoLF-PHC
algorithm can provide uncertainty in the action selection and
fools the jammer's attacks in the presence of the unknown   jamming model.

In WoLF-PHC, the mixed policy $\pi (s,a)$ is updated by increasing
the probability that it selects the most valuable action with the
highest Q-function value by a learning rate  $\xi \in (0,1]$, and
reducing other probabilities by $ - \xi /(|\mathcal{A}| - 1)$, i.e.,
\begin{equation}
\begin{split}
\begin{array}{l}
\pi (s,a) \leftarrow \pi (s,a)
 + \left\{ \begin{array}{l}
\xi ,\;{\rm{if}}\;a = \arg {\max _{a'}}Q(s,a'),\\
 - {\textstyle{\xi  \over {|\mathcal{A}| - 1}}},\;{\rm{otherwise}}{\rm{.}}
\end{array} \right.
\end{array}
\end{split}
\end{equation}

The  WoLF-PHC-based joint power allocation and reflecting
beamforming approach for the IRS-assisted communication system
against smart jamming is summarized  in \textbf{Algorithm 1}. At each episode training step, the learning agent observes
its system state  ${s^t}$ (i.e., the estimated jamming power, SINR
values, and channel coefficients) by interacting with the
environment. At each learning time slot $t$, the joint action
${a^t}$ (i.e., power allocation and reflecting beamforming) is
selected by using the probability distribution  $\pi
({s^t},{a^t})$. The $\varepsilon $-greedy policy method is
employed to balance the exploration and the exploitation,
for example, the action with the maximum Q-function value is chosen
with probability $1 - \varepsilon $ according to the known
knowledge, while a random action is chosen with probability
$\varepsilon $  based on the unknown knowledge. After executing
the selected action ${a^t}$, the environment will feedback a
 reward $r({s^t},{a^t})$ and a new system state ${s^{t +
1}}$ to the learning agent. Then, the WoLF-PHC algorithm updates both the
current policy $\pi ({s^t},{a^t})$ and updates the variable
learning rate  $\xi $ to improve the learning rate. Finally, the learning model is trained successfully,
and it can be loaded to search the joint power allocation ${\{
{P_k}\} _{k \in \mathcal{K}}}$  and reflecting beamforming matrix
${\bf{\Phi }}$ strategies according to the selected action.

\linespread{0.92}{
\begin{algorithm}[t]
\begin{normalsize}

\caption{\normalsize  WoLF-PHC-Based Joint Power Allocation
and Reflecting Beamforming}

1:$~$\textbf{Input:}  WoLF-PHC learning structure and IRS-assisted system with a jammer.\\
2:$~$\textbf{Initialize:}  $Q(s,a) = 0$, $\pi (s,a) = {1
\mathord{\left/
 {\vphantom {1 {|\mathcal{A}|}}} \right.
 \kern-\nulldelimiterspace} {|\mathcal{A}|}}$, $\xi $, $\gamma $, and $\alpha $.\\
3: $~$\textbf{for} each episode $j$  $=$ 1, 2, \dots,  ${N^{{\rm{epi}}}}$ \textbf{do}\\
4: $~~$ \textbf{for} each time step $t$ $=$ 0, 1, 2, \dots, $T$ \textbf{do}\\
5: $~~~$ Observe an initial system state  ${s^t}$;\\
6: $~~~$ Select an action $a^t$ based on the $\varepsilon $-greedy
policy: \\
$~~~~~~~~~$ ${a^t} = \arg \mathop {\max }\limits_{{a^t} \in \mathcal{A}}
Q({s^t},{a^t})$,  with probability 1-$\varepsilon $;\\
$~~~~~~~~~$ ${a^t} = {\rm{random}}{\{ {a_i}\} _{{a_i} \in
\mathcal{A}}}$, with probability  $\varepsilon $;\\
7: $~~$  Execute the exploration action ${a^t}$, receive a
reward $r({s^t},{a^t})$ and the next state ${s^{t + 1}}$;\\
8: $~~$  Update $Q({s^t},{a^t})$ by via (11);\\
8: $~~$  Update the current policy $\pi ({s^t},{a^t})$;\\
9: $~~$ \textbf{end for}\\
10: \textbf{end for}\\
11:  \textbf{Return:}  WoLF-PHC-based learning model;\\
12: \textbf{Output:} Load the learning model to achieve the joint power
allocation ${\{ {P_k}\} _{k \in \mathcal{K}}}$  and reflecting
beamforming matrix ${\bf{\Phi }}$ strategy.\\
\end{normalsize}
\label{alg_lirnn}
\end{algorithm}
}

\section{Simulation Results and Analysis}

\begin{figure}
\centering
\includegraphics[width=0.8\columnwidth]{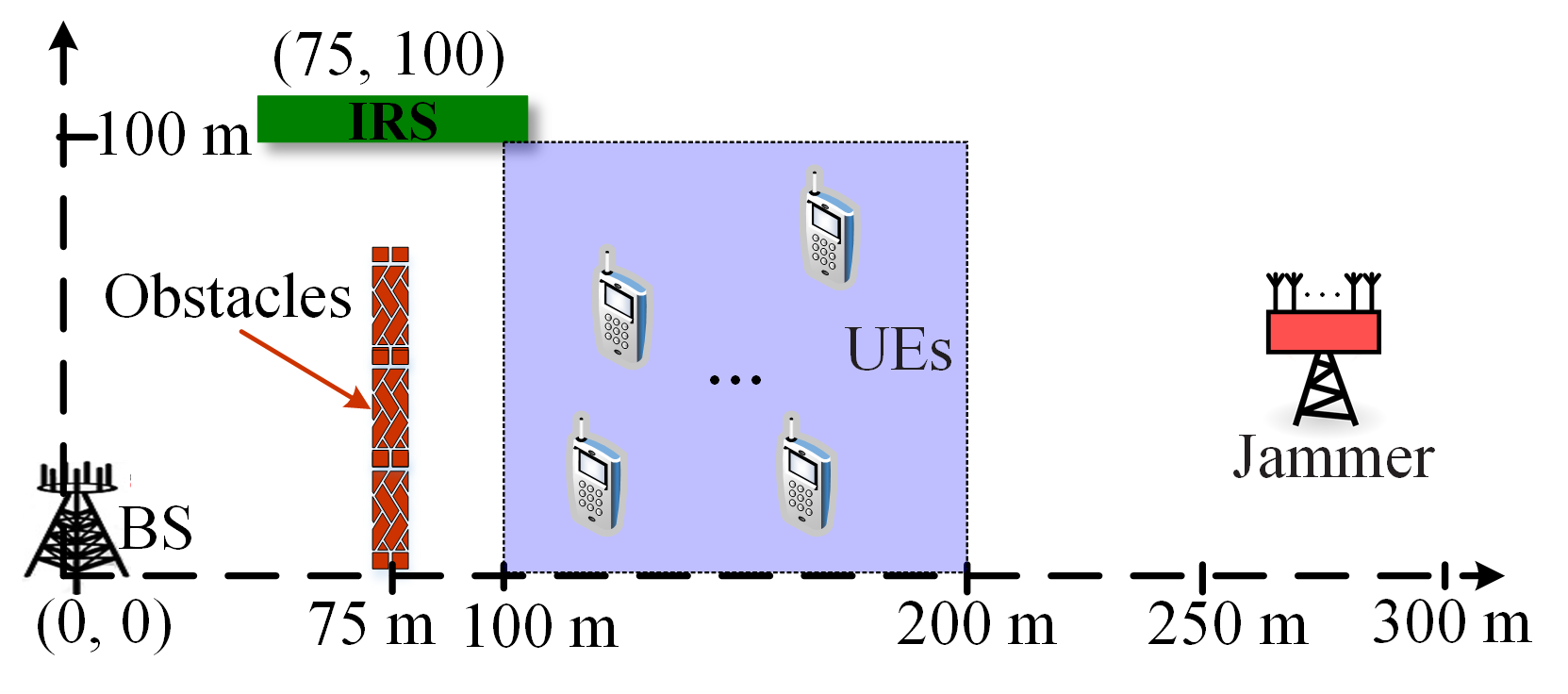}
\caption{{Simulation setup.} } \label{fig:Schematic}
\end{figure}

 This section evaluates the
performance of the IRS-assisted communication system against smart
jamming shown in Fig. 3, where a number of single-antenna UEs are
randomly located in the 100 m $\times$ 100 m right-hand side
rectangular area (light blue area). The locations of the BS,  
the IRS, and the jammer are (0, 0), (75, 100),  and (0, 0) in meter (m), respectively. There
exists obstacles which block the direct communication links
from the BS to the UEs, so the obstacles cause the large-scale pathloss for the communication links.

As for the communication channel coefficients, the path loss in dB
is expressed as
\begin{equation}
\begin{split}
PL = \left( {P{L_0} - 10\beta \; {{\log }_{10}}(d/{d_0})} \right)
\end{split}
\end{equation}
where $P{L_0}$ denotes the path loss at the reference distance
${d_0}$,  $\beta $ is the path loss exponent, and $d$ is the
distance from the transmitter to the receiver, respectively. Here,
we use ${\beta _{{\rm{bu}}}}$,  ${\beta _{{\rm{br}}}}$, ${\beta
_{{\rm{ru}}}}$, and ${\beta _{{\rm{ju}}}}$ to denote the path loss
exponents of the channel links between the BS and the UEs, between the BS
and the IRS, between the IRS and the UEs, and between the jammer and the
UEs, respectively. According to [18]-[22], we set $P{L_0} =
30\;{\rm{dB}}$,  ${d_0} = 1\;{\rm{m}}$, ${\beta
_{{\rm{bu}}}} = 3.75$, ${\beta _{{\rm{br}}}} =
{\beta _{{\rm{ru}}}} = 2.2$ and ${\beta _{{\rm{ju}}}} = 2.5$.  We set that the background noise at all
UEs is equal to  ${\delta ^2} =  - 100$ dBm. The number of
antennas at the BS and the jammer are set to $N = {N_{\rm{J}}} =
8$.  The maximum transmit power ${P_{\max }}$  at the
BS varies from 15 dBm to 40 dBm, and the number of IRS elements
$M$ varies from 20 to 100 for different simulation settings. In
addition, the jamming power of the smart jammer ranges from 15 dBm
to 40 dBm according to its jamming behavior, and the BS cannot
know the current jamming power levels, but it can
estimate the previous jamming power levels according to the
historical channel quality. The learning rate is set to  $\alpha  = 5 \times 10^{-3}$, the discount factor
is set to $\gamma  = 0.9$ and the exploration rate is set to
$\varepsilon  = 0.1$. The cost parameter  ${\lambda _1}$ in (7) is set to ${\lambda _1} = 1$. We set ${\xi} = 0.04$ [23], [24]. 

In addition, we compare the proposed WoLF-PHC-based
joint power allocation and reflecting beamforming approach
(denoted as WoLF-PHC learning) with the following
approaches:

\begin{itemize}
\item The popular fast Q-Learning approach [13], which is adopted to optimize the transmit power allocation and reflecting beamforming in IRS-assisted communication systems (denoted as fast Q-Learning [13]).

\item The greedy approach which jointly optimizes the
BS's transmit power allocation and the IRS's reflect beamforming (denoted as Greedy).

\item  The optimal transmit power allocation at the BS without IRS
assistance (denoted as Optimal PA without IRS).
\end{itemize}

We first compare the convergence performance of all approaches when
${P_{\max }} = 30\;{\rm{dBm}}$,  $K=4$, and $M = 60$. It is observed that the system
 rate of all approaches (except the
optimal PA approach) increases with the number of iterations, and the proposed  WoLF-PHC  learning approach
accelerates the convergence rate and enhances the system rate
 compared with both the
fast Q-Learning approach and the greedy approach. Because the
proposed leaning approach adopts WoLF-PHC to
increase the learning rate and enhance the learning
efficiency against smart jamming, yielding a faster learning rate
under the dynamic environment. Among all approaches, the fast Q-Learning
requires the largest number of convergence iterations to optimize
the Q-function estimator, where the slow convergence may fail to
protect anti-jamming performance against smart jamming in real-time
systems. Moreover, the optimal PA approach
without IRS has the fastest convergence speed, but it obtains the  worst performance among all approaches, because it does not employ an
IRS for system performance improvement and jamming resistance.

\begin{figure}
\centering
\includegraphics[width=0.8\columnwidth]{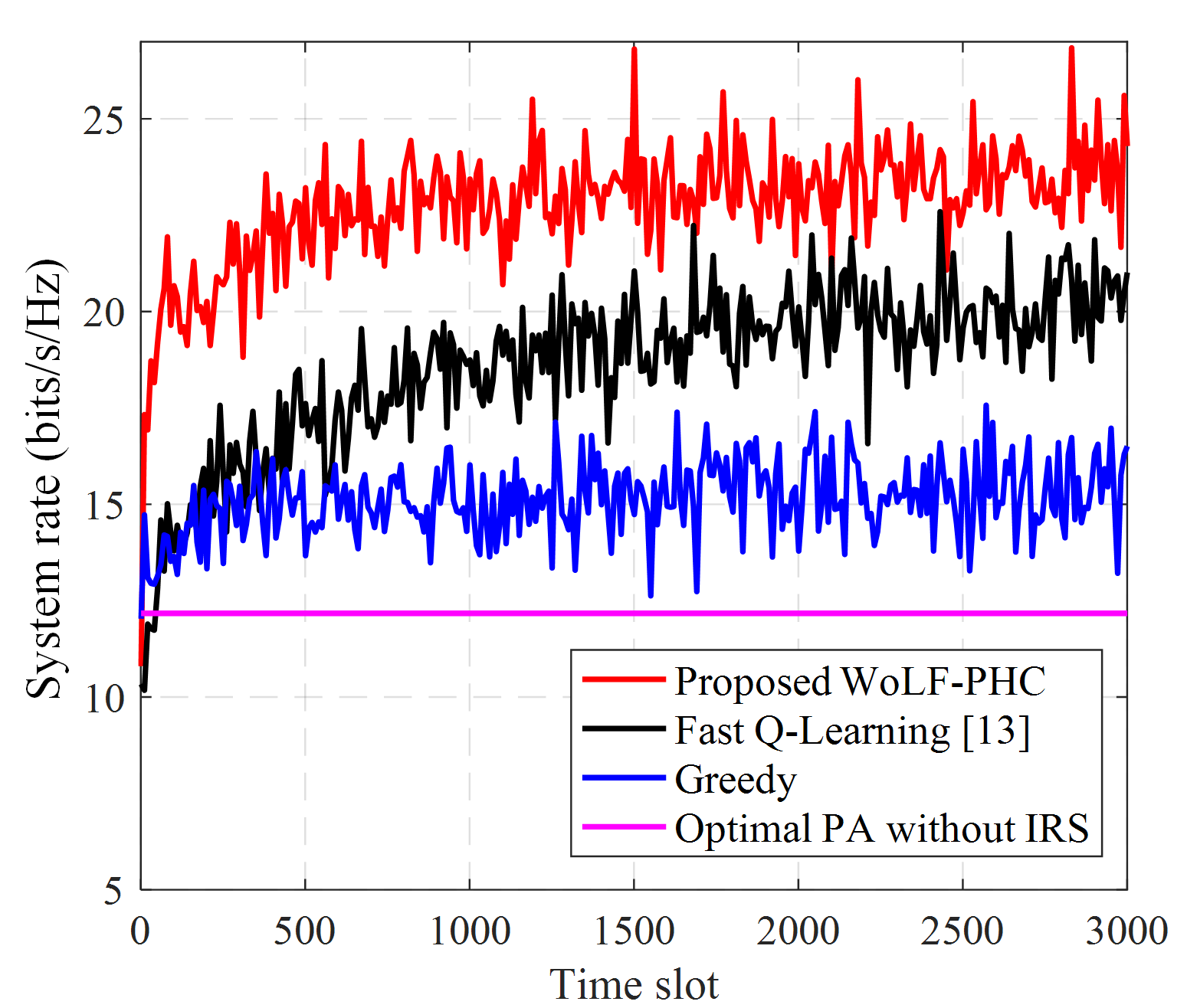}
\caption{{Convergence behaviors of the various approaches.} }
\label{fig:Schematic}
\end{figure}

\begin{figure}
\centering
\includegraphics[width=0.8\columnwidth]{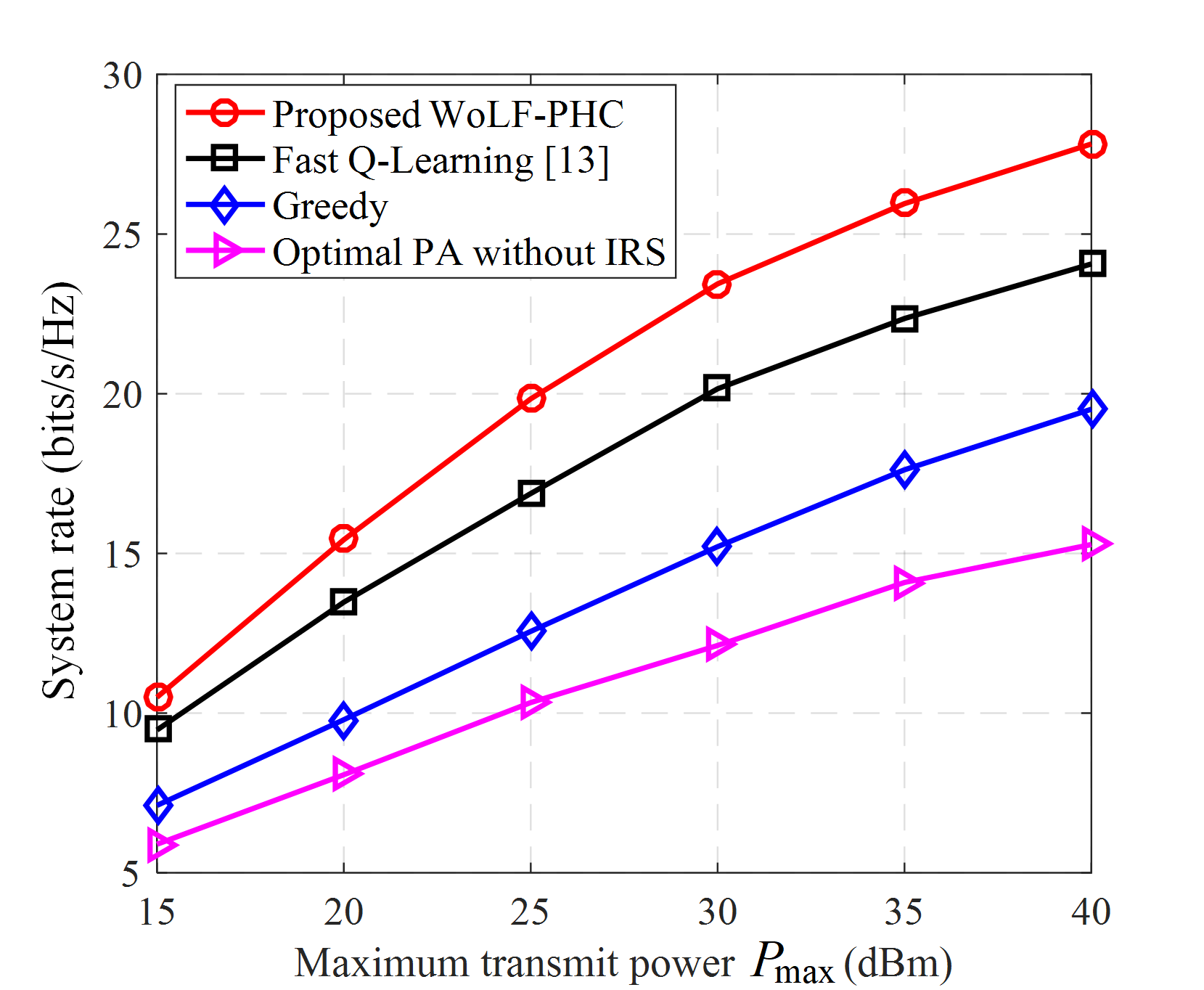}
\caption{{Performance comparisons versus the maximum transmit
power ${P_{\max }}$.} } \label{fig:Schematic}
\end{figure}

The average system rate  versus the
maximum transmit power ${P_{\max }}$  for various approaches are
shown in Fig. 5 when  $K = 4$, and $M = 60$, which demonstrates that the achieved system rate improve as  ${P_{\max }}$ increases. We can also
observe that both the proposed learning approach and the fast Q-Learning approach have good system rate value under different values of  ${P_{\max }}$, and both of them greatly outperform other
approaches.  Additionally, the performance improvement achieved by using IRS versus without IRS increases  with  ${P_{\max }}$, which indicates the advantage of
deploying the IRS against smart jamming. In addition, the
performance of both the system rate 
of the proposed WoLF-PHC-based learning approach is higher 
than that of the fast Q-Learning approach, which  is due to the fact
that WoLF-PHC is adopted to effectively search the
optimal joint power allocation and reflecting beamforming strategy
against smart jamming in dynamic uncertain environments.

\begin{figure}
\centering
\includegraphics[width=0.8\columnwidth]{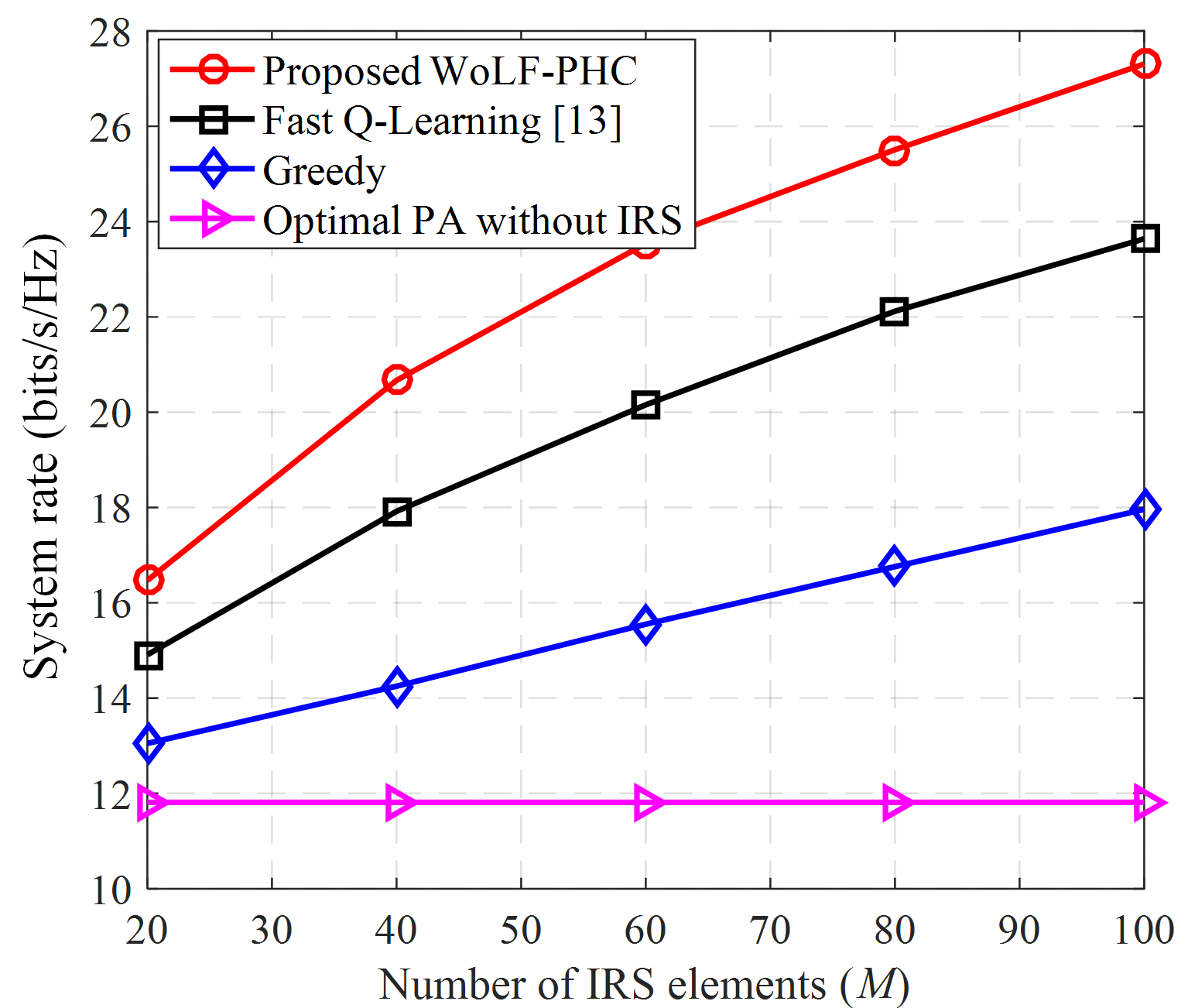}
\caption{{Performance comparisons versus the number of IRS
elements.}}  \label{fig:Schematic}
\vspace{-5pt} 
\end{figure}

Fig. 6 compares the performance of the four approaches with the
different reflecting elements number $M$ when ${P_{\max }} =
30\;{\rm{dBm}}$ and   $K = 4$.
It can be seen that except the optimal PA approach without IRS, the
performance of all IRS-based approaches increases with $M$,
and greatly outperforms the optimal PA approach without IRS. This is that the IRS has the ability to support higher degrees of freedom for
performance optimization, resulting in the great performance
gains obtained by employing the IRS against smart jamming over the
traditional system without IRS. Specifically, when  $M = 20$, the
system achievable rate gain of the proposed learning approach over
the optimal PA approach without IRS is only about 4.21 bits/s/Hz,
while this value is improved to 15.47 bits/s/Hz when  $M
= 100$. Such performance improvement results
from the facts that the more power can be achieved at the IRS by
increasing $M$, and the higher reflecting beamforming gain is
achieved to design the IRS phase shifts to improve the received
desired signal as well as mitigate the jamming interference from
the smart jammer by increasing $M$.

\vspace{-5pt} 
 \section{Conclusions}

This paper proposed to improve the anti-jamming performance
of wireless communication systems by employing an IRS. Specifically, we formulated an
optimization problem by joint optimizing both the transmit power
allocation at the BS and the reflecting beamforming at the IRS. A  WoLF-PHC learning approach was
proposed to achieve the optimal anti-jamming 
strategy, where WoLF-PHC is capable of quickly achieving
the optimal policy without knowing the jamming model. Simulation results confirmed that the IRS-assisted
system significantly improves the anti-jamming performance compared with other approaches. We will pay attention to apply IRS in visible light communication systems in the future [25], [26].

\end{document}